\pdfoutput=1
\documentclass[a4paper,11pt]{article}
\usepackage{pos}

\renewcommand{\speakerText}{Speaker}

\graphicspath{{images/}}

\usepackage{float}
\usepackage[section]{placeins}
\usepackage{needspace}

\usepackage{orcidlink}
\newcommand{\orcid}[1]{\,\orcidlink{#1}}
\let\posPrintHeadAuthors\printHeadAuthors
\renewcommand{\printHeadAuthors}{%
  \begingroup\renewcommand{\orcid}[1]{}\posPrintHeadAuthors\endgroup}

\title{Point-cloud generative models for fast calorimeter simulation across particles and geometries}
\ShortTitle{Point-cloud generative models for fast calorimeter simulation}

\author[a,b]{Thorsten Buss\orcid{0000-0002-1717-2138}}
\author[b]{Henry Day-Hall\orcid{0000-0002-9710-2980}}
\author[b]{Frank Gaede\orcid{0000-0002-7055-9200}}
\author[a]{Gregor Kasieczka\orcid{0000-0003-3457-2755}}
\author[b]{Katja Kr\"uger\orcid{0000-0002-1956-6608}}
\author[b]{Anatolii Korol\orcid{0000-0002-2569-1771}}
\author[b]{Thomas Madlener\orcid{0000-0002-0128-6536}}
\author[c]{Peter McKeown\orcid{0009-0006-9722-2233}}
\author[a]{Martina Mozzanica\orcid{0009-0002-1111-6247}}
\author*[a]{Lorenzo Valente\orcid{0009-0007-0080-8738}}

\affiliation[a]{Universit\"at Hamburg, Institut f\"ur Experimentalphysik,\\
Luruper Chaussee 149, 22761 Hamburg, Germany}

\affiliation[b]{Deutsches Elektronen-Synchrotron DESY,\\
Notkestra\ss e 85, 22607 Hamburg, Germany}

\affiliation[c]{CERN, 1211 Geneva 23, Switzerland}

\emailAdd{lorenzo.valente@uni-hamburg.de}

\abstract{%
Detailed \texttt{Geant4} simulation of calorimeter showers is the largest single computing
cost of collider experiments, and the High-Luminosity LHC will need about ten times more
simulated events than are currently produced. We summarise recent
progress in generative point cloud fast simulation for highly granular calorimeters.
\textsc{CaloClouds3} generates photon (electromagnetic) showers, is geometry-independent,
and runs on average about $120\times$ faster than \texttt{Geant4} on a single CPU. \textsc{CaloHadronic} uses transformer attention
to extend the point cloud diffusion approach to pion (hadronic) showers spanning the
electromagnetic and hadronic calorimeters. \textsc{AllShowers} unifies twelve particle
types in a single model with far fewer parameters than the specialised baselines while
matching or exceeding their fidelity. We close with cross-geometry transfer learning,
which needs two to three orders of magnitude fewer training showers while preserving the
generative performance.%
}

\FullConference{14th Large Hadron Collider Physics Conference (LHCP2026)\\
18--22 May 2026\\
Paris, France\\}

\begin{document}
\maketitle

% =====================================================================
%  main.tex = orchestrator. Section content lives in chapters/.
%
%  Page limits (PoS, excluding first pages and references):
%    - plenary talks:            8 pages
%    - parallel talks / posters: 4 pages
%  Target for THIS contribution: 4 pages (excluding first page and references).
%  Submission deadline: 25 September 2026.
% =====================================================================

\section{Introduction}
Detailed detector simulation is the largest consumer of computing resources at the LHC,
and calorimeter showers account for most of its cost. A single event simulated with
\texttt{Geant4}~\cite{GEANT4:2002zbu} costs on the order of $10^2$ to $10^4$ CPU-seconds,
and analyses require roughly ten simulated events for every recorded one. The
High-Luminosity LHC will need about ten times more simulated events.

Deep generative models could reproduce the energy deposits orders of magnitude faster than
\texttt{Geant4}. They fall into two broad families depending on their native shower
representation: fixed-grid models, tied to the segmentation of one particular detector,
and point cloud models, which describe a shower as a set of energy deposits whose number
varies from shower to shower~\cite{Buhmann:2023bwk,Buhmann:2023kdg}. A point cloud
presupposes no readout segmentation, so it can describe detectors with complex geometries,
although the deposits themselves are still shaped by the detector that produces the
shower. We therefore adopt a point cloud representation. The models below assume a
sampling calorimeter, whose active and passive layers set the layer structure of the
point cloud.

This contribution reviews four lines of work in point cloud generative fast
simulation, from a single particle type in one detector
(\textsc{CaloClouds3}~\cite{Buss:2025kiu} for photons,
\textsc{CaloHadronic}~\cite{Buss:2025cyw} for hadronic pions) to twelve particle types in
one compact model (\textsc{AllShowers}~\cite{Buss:2026yrf}) and to new geometries through
transfer learning~\cite{Gaede:2025shc,Buss:2026tlx}.

\section{CaloClouds3: fast photon showers}
\textsc{CaloClouds3}~\cite{Buss:2025kiu} is the latest
iteration of the CaloClouds series~\cite{Buhmann:2023bwk,Buhmann:2023kdg}, trained on
photon showers in the highly granular electromagnetic calorimeter of ILD. As for all
models below, the photons enter at the calorimeter surface, so any interaction in the
material in front of it is left to \texttt{Geant4}. A normalising
flow, ShowerFlow, predicts the number of points and the visible energy in each layer, and
a single-step distilled diffusion model then generates the individual points.
Conditioning on the incident direction and a training procedure independent of the impact
position let one model cover the whole detector barrel, whereas the predecessor
model~\cite{Buhmann:2023kdg} was restricted to a single fixed incident angle and position.

\begin{figure}[!htb]
  \centering
  % Caption beside the plot (review round, Sept 2026): saves vertical space
  % without shrinking the plot.
  \begin{minipage}[c]{0.62\linewidth}
    \centering
    % Upper panel of the two-panel timing figure (time per event), followed by the
    % x-axis strip of the lower panel (tick labels + label). Both crops share the
    % same left/right trim, so the tick labels line up with the upper panel.
    \includegraphics[width=\linewidth,trim=8 163 10 8,clip]{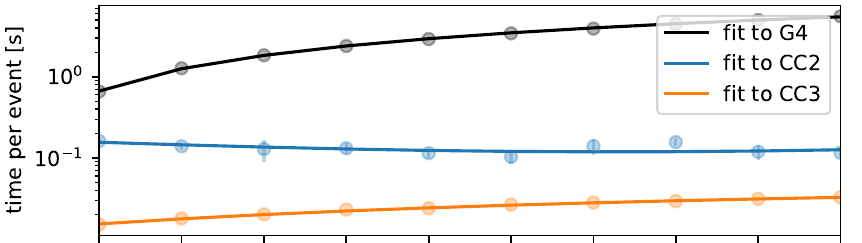}\par\nointerlineskip
    \includegraphics[width=\linewidth,trim=8 10 10 250,clip]{caloclouds3_timing}
  \end{minipage}\hfill
  \begin{minipage}[c]{0.35\linewidth}
    \caption{Simulation time per event of \textsc{CaloClouds3}, \textsc{CaloClouds~II} and
    \texttt{Geant4} on a single CPU thread as a function of the incident photon energy.
    Adapted from Ref.~\cite{Buss:2025kiu}.}
    \label{fig:caloclouds3}
  \end{minipage}
\end{figure}

On a single CPU, \textsc{CaloClouds3} is on average about $120\times$ faster than
\texttt{Geant4} (Fig.~\ref{fig:caloclouds3}), and the gain grows with the incident
energy, from roughly $45\times$ at $10$~GeV to $130\times$ at $50$~GeV and $180\times$
at $100$~GeV. The one-off overhead is the simulation of $3\times10^6$ \texttt{Geant4}
training showers and the training of the model. The
model reproduces the shower observables and, after full reconstruction, the di-photon
separation, in agreement with \texttt{Geant4} within the expected fluctuations.

\section{CaloHadronic: hadronic showers}
Hadronic showers are the dominant contribution to the simulation cost.
\textsc{CaloHadronic}~\cite{Buss:2025cyw} is the first model to generate charged pion
($\pi^+$) showers holistically across both the electromagnetic (ECal) and hadronic (HCal)
calorimeters of a highly granular detector. Hadronic showers have a track-like
substructure, a larger spatial extent and larger fluctuations, so the point-wise layers
used for the more isotropic electromagnetic showers are replaced by transformer
self-attention, which captures the structure between points. The three-stage pipeline
chains a point-count model that predicts the number of points in each layer, a diffusion model for the
ECal points and a second diffusion model that adds the HCal points conditioned on the
ECal points of the same shower generated in the previous stage.

\begin{figure}[!htb]
  \centering
  \includegraphics[width=0.335\linewidth]{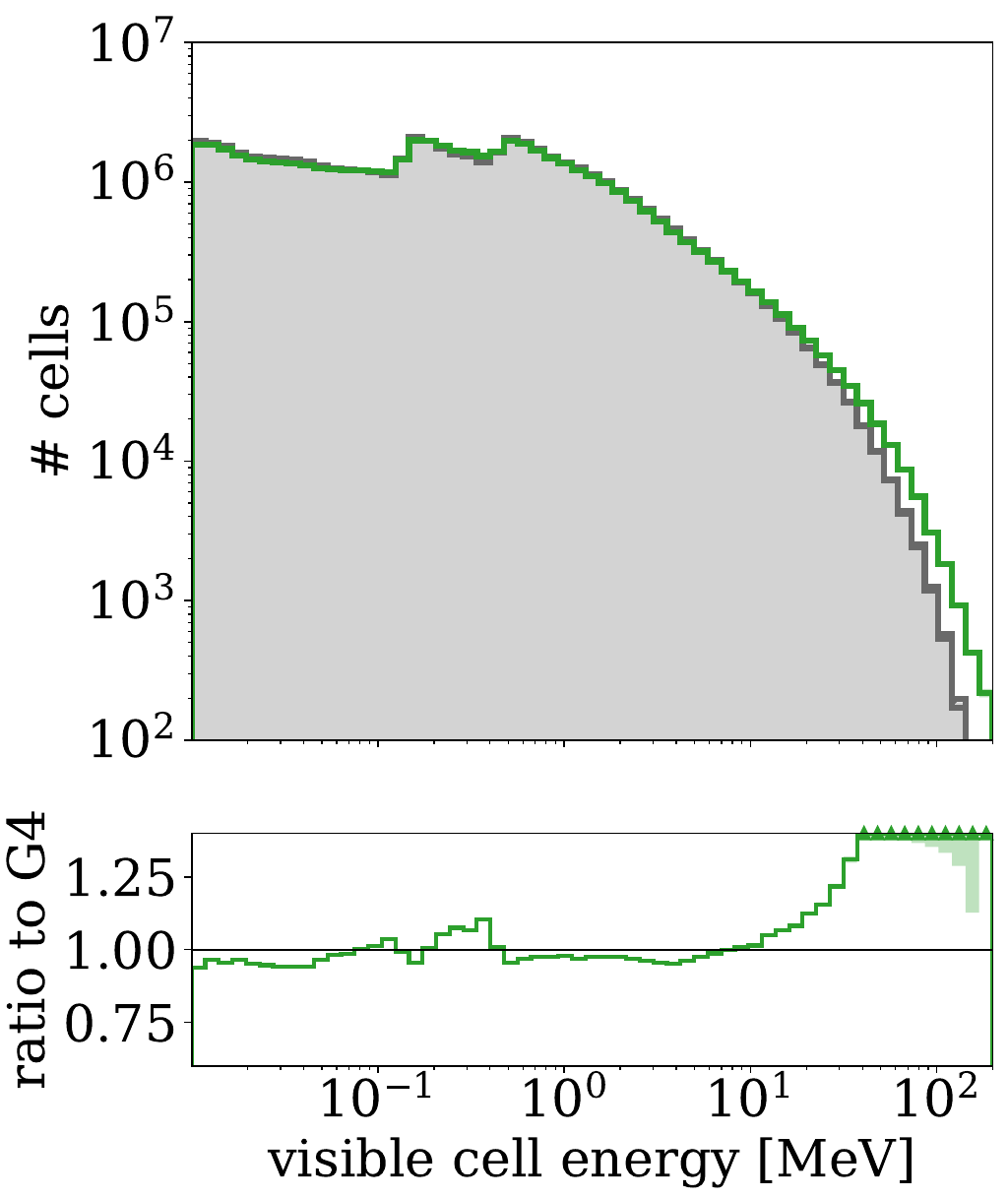}\hfill
  \includegraphics[width=0.315\linewidth]{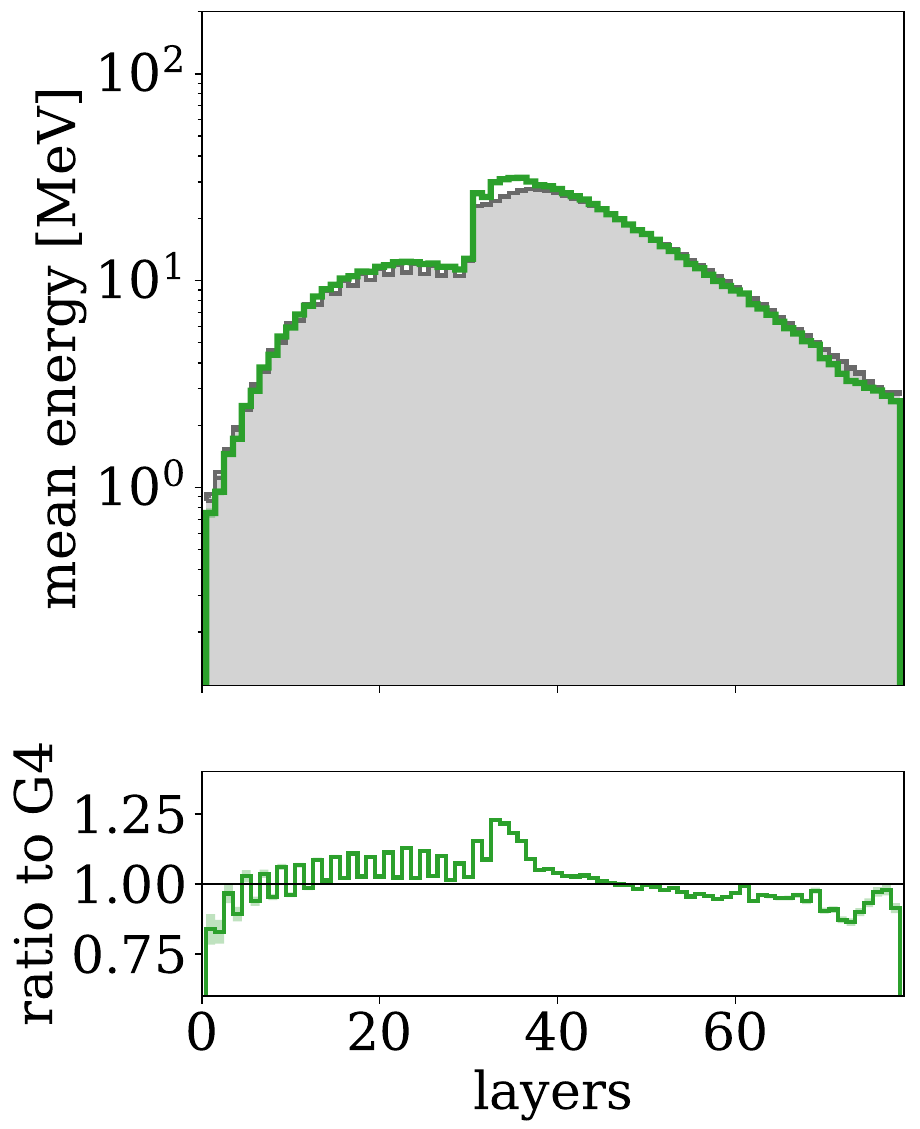}\hfill
  \includegraphics[width=0.31\linewidth]{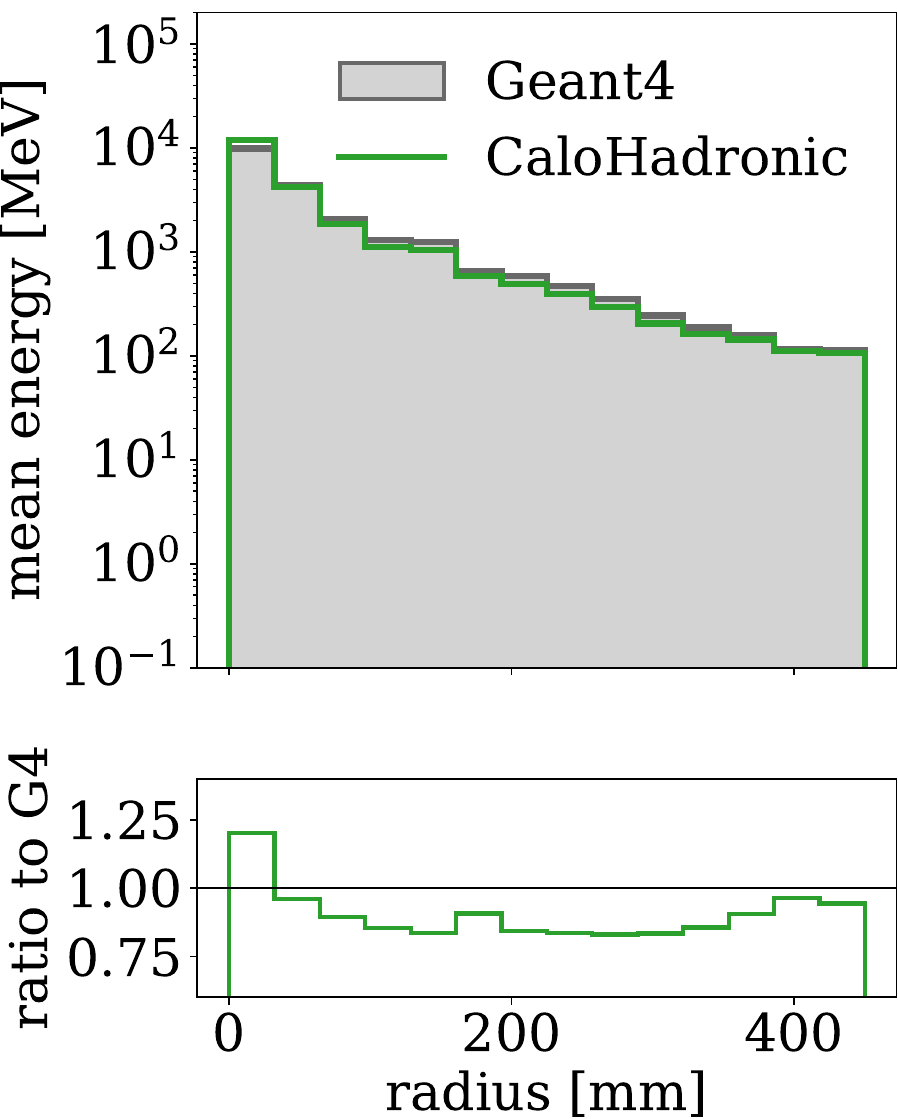}
  \caption{\textsc{CaloHadronic} pion showers compared with \texttt{Geant4}: cell energy spectrum
  (left), longitudinal shower profile (centre) and radial shower profile (right), with
  the ratio to \texttt{Geant4} below each panel. Distributions from 50\,000 showers with
  incident energies uniform between 10 and 90~GeV, error bands are the statistical
  uncertainty per bin. Adapted from Ref.~\cite{Buss:2025cyw}.}
  \label{fig:calohadronic}
\end{figure}

Evaluated on pion showers with incident energies between 10 and 90~GeV, the model
reproduces the cell energy spectrum, including the minimum ionising particle (MIP) peaks
of both calorimeters, and the longitudinal and radial shower profiles
(Fig.~\ref{fig:calohadronic}). The agreement also holds after full
\texttt{PandoraPFA} particle flow reconstruction, in the number, energy, momentum and
type of the particle flow objects. On a GPU the speed-up over \texttt{Geant4} reaches
three orders of magnitude in the best case and ranges from a few times to about
$15\times$ at the sampling setting used, although the generative fidelity is still being
improved. \textsc{CaloHadronic}, like \textsc{CaloClouds3}, is specialised to one particle
type, so covering all species this way would need one model per particle, which
motivates a single model for all of them.

\section{AllShowers: one model for all particles}
\textsc{AllShowers}~\cite{Buss:2026yrf} is a single generative model covering twelve
particle types across both the electromagnetic and hadronic calorimeters, namely
$e^\pm$, $\gamma$, $\pi^\pm$, $K^\pm$, $K^0_L$, $p$, $\bar{p}$, $n$ and $\bar{n}$, where
earlier models each covered only photons, electrons or $\pi^+$.
\textsc{PointCountFM}, inherited from \textsc{CaloHadronic}, predicts the number of points in each layer, and a conditional flow matching Transformer with learned embeddings of
particle type and layer generates the points.

\begin{figure}[!htb]
  \centering
  % Caption beside the plot (review round, Sept 2026): saves vertical space
  % without shrinking the plot.
  \begin{minipage}[c]{0.65\linewidth}
    \centering
    \includegraphics[width=\linewidth]{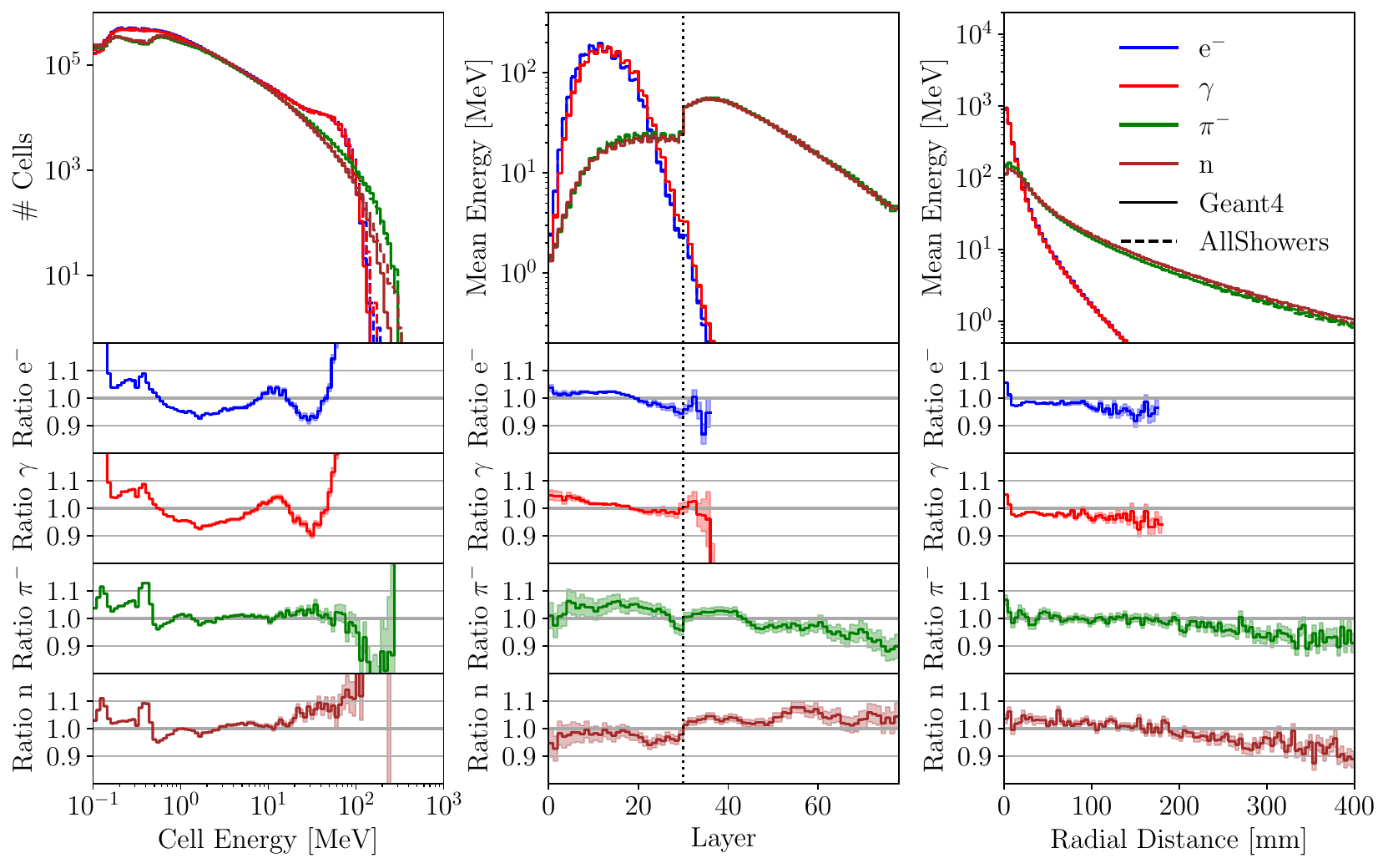}
  \end{minipage}\hfill
  \begin{minipage}[c]{0.33\linewidth}
    \caption{\textsc{AllShowers} against \texttt{Geant4} for different incident particle types
    and angles at a fixed incident energy of 100~GeV: cell energy spectrum (left),
    longitudinal (centre) and radial (right) energy distributions, with the ratio to
    \texttt{Geant4} below each panel. Adapted from Ref.~\cite{Buss:2026yrf}.}
    \label{fig:allshowers}
  \end{minipage}
\end{figure}

With only ${\sim}0.6$M trainable parameters, \textsc{AllShowers} replaces the
specialised baselines \textsc{CaloClouds3}~\cite{Buss:2025kiu} (${\sim}6.1$M) and
\textsc{CaloHadronic}~\cite{Buss:2025cyw} (${\sim}2.1$M), and is trained in under a day
on 16 A100 GPUs on 2M showers of $5$ to $126$~GeV and up to ${\sim}6000$ points each.
Despite the smaller size it matches or
exceeds the fidelity of the specialised models and stays within about $10\%$ of
\texttt{Geant4} for most observables (Fig.~\ref{fig:allshowers}). Combining the
particle types lets them share their common physics behaviour. For pions the
longitudinal profile is closer to \texttt{Geant4} than in \textsc{CaloHadronic}, likely
helped by a learned layer embedding, and sampling is faster, both per function
evaluation and in their number, 32 instead of 59. For photons \textsc{CaloClouds3}
remains at least two orders of magnitude faster on CPU, as it generates the points
independently in a single distilled step. A model still has to be trained per detector,
which motivates transfer learning across geometries.

\section{Transfer learning across geometries}
A surrogate stays bound to the geometry it was trained on, so a new detector demands a
large new \texttt{Geant4} dataset and loses the speed-up. Ref.~\cite{Gaede:2025shc}
adapts a point cloud model pre-trained on ILD photon showers to the electrons of
CaloChallenge Dataset~3~\cite{Krause:2024avx}. With only $100$ target showers, full
fine-tuning reduces the geometric mean Wasserstein distance to \texttt{Geant4} by about
$44\%$ relative to training from scratch, and bias-only fine-tuning (BitFit), which updates
$17\%$ of the parameters and preserves the inference speed, stays within about $7\%$ of
full fine-tuning on the same aggregated geometric mean Wasserstein distance.

\begin{figure}[!htb]
  \centering
  \includegraphics[width=0.72\linewidth]{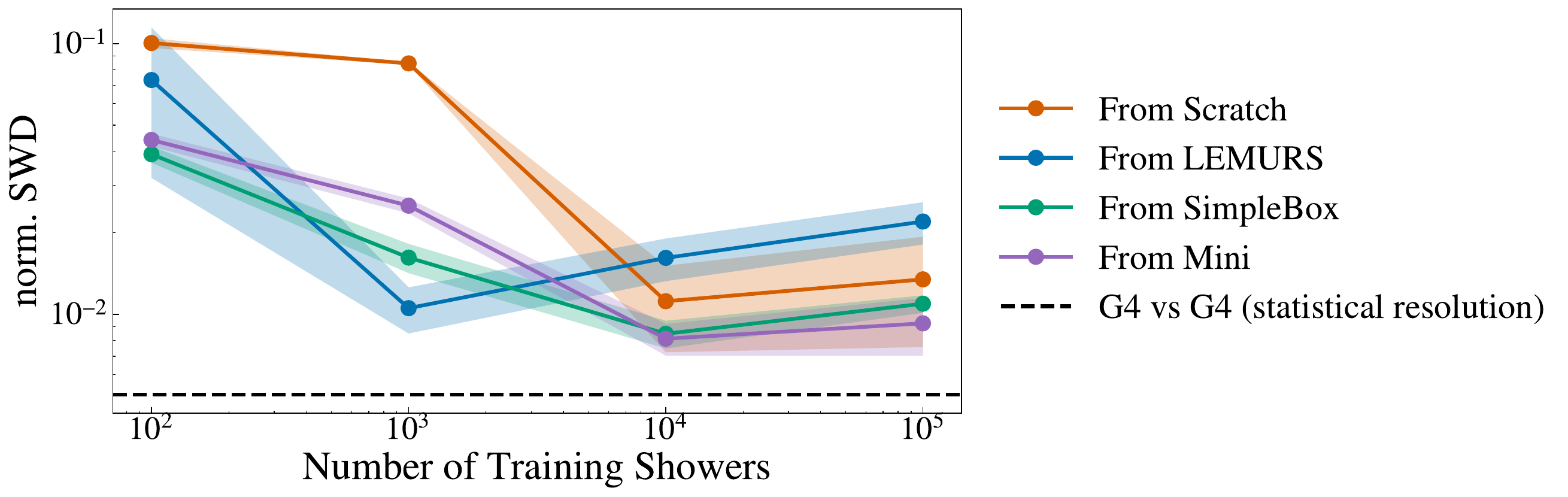}
  \caption{Sample efficiency on the held-out FCC-ee ALLEGRO calorimeter: geometric mean
  over the three observables of the normalised sliced Wasserstein distance (SWD) to
  \texttt{Geant4} as a function of the fine-tuning dataset size, for training from
  scratch and for fine-tuning from the \textsc{LEMURS}, \textsc{SimpleBox} and Mini
  pre-trainings. Bands span the mean $\pm$ one standard deviation over five seeds, the
  dashed line is the \texttt{Geant4}-versus-\texttt{Geant4} statistical resolution.
  Adapted from Ref.~\cite{Buss:2026tlx}.}
  \label{fig:transfer}
\end{figure}

Ref.~\cite{Buss:2026tlx} extends the \textsc{AllShowers}~\cite{Buss:2026yrf} backbone
with a geometry-aware conditioning on two scalars of the calorimeter design, the sampling
fraction and the number of layers, and pre-trains it on two alternative sets of detectors.
One is \textsc{SimpleBox}, a new family of $10^4$ synthetic tungsten-silicon box
calorimeters spanning both scalars (15 to 45 layers, sampling fractions $0.01$ to $0.05$),
simulated with \texttt{Geant4} and publicly released~\cite{mgg_datasets}.
The other is four of
the five realistic \textsc{LEMURS} calorimeters~\cite{McKeown:2025gtw}, all silicon or
scintillator sampling calorimeters with tungsten or lead absorbers (40 to 90 layers,
sampling fractions $0.026$ to $0.033$), re-simulated as
point clouds on a $1\,\mathrm{mm}\times1\,\mathrm{mm}$ grid instead of
the cell-aggregated Universal Grid Representation of the original release. Both priors are trained at a matched $4\times10^6$-shower budget, and the synthetic
prior transfers to all five \textsc{LEMURS} detectors, unseen in its pre-training.
The test is the fifth \textsc{LEMURS} calorimeter, FCC-ee ALLEGRO, a noble-liquid detector
(11 lead and liquid-argon layers) left out of both pre-trainings, whose sampling fraction
of $0.16$ lies outside the pre-training region. Fine-tuned on only $10^3$ ALLEGRO showers, the two priors cut the aggregated sliced
Wasserstein distance to \texttt{Geant4} (Fig.~\ref{fig:transfer}) by factors of $5.2$
(\textsc{SimpleBox}) and $8.0$ (\textsc{LEMURS}) over training from scratch on the same
showers. Beyond a thousand target showers, the synthetic prior overtakes the realistic one,
which degrades, and a Mini pre-training on $2.5\%$ of the \textsc{SimpleBox} pool still
beats training from scratch at every fine-tuning size. Fine-tuning matches the
from-scratch agreement with a hundred times fewer target showers, so geometric diversity
alone, with no realistic detector involved, is a practical way to pre-train a
transferable shower model. The prior so far covers only photon showers and two geometry
scalars, and extending it is the natural next direction.

\Needspace*{12\baselineskip}
\section{Conclusions}
The four lines of work share the point cloud representation and together remove the
main limitations of generative fast calorimeter simulation, from geometry-independent
photon showers~\cite{Buss:2025kiu} and holistic hadronic showers~\cite{Buss:2025cyw} to
twelve particle types in one compact model~\cite{Buss:2026yrf}.
Transfer learning~\cite{Gaede:2025shc,Buss:2026tlx} reaches a new geometry from very few
target showers and cuts the simulated data needed to train the network by two to three
orders of magnitude. Open directions
are to improve the hadronic fidelity and to extend the synthetic geometry prior to all
particle species and to the full calorimeter description, towards zero-shot simulation of
unseen detectors.

\section*{Acknowledgements}
We thank Dirk Zerwas for the careful review of the manuscript.
This work used the Maxwell computational resources at DESY. It has received funding from
the European Union's Horizon 2020 Research and Innovation programme under Grant Agreement
No 101004761, from the Deutsche Forschungsgemeinschaft under Germany's Excellence
Strategy, EXC 2121 Quantum Universe, 390833306, and via the KISS consortium (05D23GU4,
13D22CH5) funded by the German Federal Ministry of Research, Technology and Space (BMFTR)
in the ErUM-Data action plan. A.K. is supported by the Helmholtz Initiative and Networking
Fund's initiative for refugees as a refugee of the war in Ukraine. P.M. is supported by
the CERN Strategic R\&D Programme on Technologies for Future Experiments~\cite{EPRD}.

% Bibliography (BibTeX): biblio.bib + JHEP.bst auto-link DOI/arXiv.
\bibliographystyle{JHEP}
\bibliography{biblio}

\providecommand{\href}[2]{#2}\begingroup\raggedright\begin{thebibliography}{10}

\bibitem{GEANT4:2002zbu}
{\scshape GEANT4} collaboration, \emph{{GEANT4 - A Simulation Toolkit}},
  \href{https://doi.org/10.1016/S0168-9002(03)01368-8}{\emph{Nucl. Instrum.
  Meth. A} {\bfseries 506} (2003) 250}.

\bibitem{Buhmann:2023bwk}
E.~Buhmann, S.~Diefenbacher, E.~Eren, F.~Gaede, G.~Kasieczka, A.~Korol et~al.,
  \emph{{CaloClouds: fast geometry-independent highly-granular calorimeter
  simulation}},
  \href{https://doi.org/10.1088/1748-0221/18/11/P11025}{\emph{JINST} {\bfseries
  18} (2023) P11025} [\href{https://arxiv.org/abs/2305.04847}{{\ttfamily
  2305.04847}}].

\bibitem{Buhmann:2023kdg}
E.~Buhmann, F.~Gaede, G.~Kasieczka, A.~Korol, W.~Korcari, K.~Kr{\"u}ger et~al.,
  \emph{{CaloClouds II: ultra-fast geometry-independent highly-granular
  calorimeter simulation}},
  \href{https://doi.org/10.1088/1748-0221/19/04/P04020}{\emph{JINST} {\bfseries
  19} (2024) P04020} [\href{https://arxiv.org/abs/2309.05704}{{\ttfamily
  2309.05704}}].

\bibitem{Buss:2025kiu}
T.~Buss, H.~Day-Hall, F.~Gaede, G.~Kasieczka, K.~Kr{\"u}ger, A.~Korol et~al.,
  \emph{{CaloClouds3: Ultra-fast geometry-independent highly-granular
  calorimeter simulation}},
  \href{https://doi.org/10.1088/1748-0221/21/03/P03018}{\emph{JINST} {\bfseries
  21} (2026) P03018} [\href{https://arxiv.org/abs/2511.01460}{{\ttfamily
  2511.01460}}].

\bibitem{Buss:2025cyw}
T.~Buss, F.~Gaede, G.~Kasieczka, A.~Korol, K.~Kr{\"u}ger, P.~McKeown et~al.,
  \emph{{CaloHadronic: a diffusion model for the generation of hadronic
  showers}}, \href{https://doi.org/10.1088/1748-0221/21/01/P01042}{\emph{JINST}
  {\bfseries 21} (2026) P01042}
  [\href{https://arxiv.org/abs/2506.21720}{{\ttfamily 2506.21720}}].

\bibitem{Buss:2026yrf}
T.~Buss, H.~Day-Hall, F.~Gaede, G.~Kasieczka and K.~Kr{\"u}ger,
  \emph{{AllShowers: One model for all calorimeter showers}},
  \href{https://arxiv.org/abs/2601.11716}{{\ttfamily 2601.11716}}.

\bibitem{Gaede:2025shc}
F.~Gaede, G.~Kasieczka and L.~Valente, \emph{{Cross-geometry transfer learning
  in fast electromagnetic shower simulation}},
  \href{https://doi.org/10.1088/1748-0221/21/07/P07037}{\emph{JINST} {\bfseries
  21} (2026) P07037} [\href{https://arxiv.org/abs/2512.00187}{{\ttfamily
  2512.00187}}].

\bibitem{Buss:2026tlx}
T.~Buss, H.~Day-Hall, F.~Gaede, G.~Kasieczka, K.~Kr{\"u}ger, P.~McKeown et~al.,
  \emph{{Transferable Fast Calorimeter Shower Generation via Multi-Geometry
  Pre-training}},  \href{https://arxiv.org/abs/2608.18233}{{\ttfamily
  2608.18233}}.

\bibitem{Krause:2024avx}
O.~Amram et~al., \emph{{CaloChallenge 2022: a community challenge for fast
  calorimeter simulation}},
  \href{https://doi.org/10.1088/1361-6633/ae1304}{\emph{Rept. Prog. Phys.}
  {\bfseries 88} (2025) 116201}
  [\href{https://arxiv.org/abs/2410.21611}{{\ttfamily 2410.21611}}].

\bibitem{mgg_datasets}
L.~Valente et~al., ``{Point cloud calorimeter shower datasets for
  multi-geometry pre-training: SimpleBox and LEMURS}.'' Universit{\"a}t Hamburg
  Research Data Repository,
  \href{https://doi.org/10.25592/uhhfdm.19103}{\texttt{doi:10.25592/uhhfdm.19103}},
  2026.

\bibitem{McKeown:2025gtw}
P.~McKeown, P.~Raikwar and A.~Zaborowska, \emph{{LEMURS dataset: Large-scale
  multi-detector ElectroMagnetic Universal Representation of Showers}},
  \href{https://arxiv.org/abs/2509.05108}{{\ttfamily 2509.05108}}.

\bibitem{EPRD}
C.~Joram et~al., \emph{{Extension of the R\&D Programme on Technologies for
  Future Experiments}},  Tech. Rep.
  \href{https://cds.cern.ch/record/2850809}{CERN-EP-RDET-2023-001}, CERN
  (2023).

\end{thebibliography}\endgroup

\end{document}